\documentclass[prl,reprint,longbibliography] {revtex4-1}

\usepackage{amsmath}

\usepackage[utf8]{inputenc}
\DeclareUnicodeCharacter{03B4}{$\delta$}
\DeclareUnicodeCharacter{2009}{\,}
\DeclareUnicodeCharacter{03BC}{\mbox{$\mu$}}

\usepackage{times}

\usepackage[english]{babel}

\makeatletter
\def\bbl@set@language#1{%
	\edef\languagename{%
		\ifnum\escapechar=\expandafter`\string#1\@empty
		\else\string#1\@empty\fi}%
	\@ifundefined{babel@language@alias@\languagename}{}{%
		\edef\languagename{\@nameuse{babel@language@alias@\languagename}}%
	}%
	\select@language{\languagename}%
	\expandafter\ifx\csname date\languagename\endcsname\relax\else
	\if@filesw
	\protected@write\@auxout{}{\string\select@language{\languagename}}%
	\bbl@for\bbl@tempa\BabelContentsFiles{%
		\addtocontents{\bbl@tempa}{\xstring\select@language{\languagename}}}%
	\bbl@usehooks{write}{}%
	\fi
	\fi}
\newcommand{\DeclareLanguageAlias}[2]{%
	\global\@namedef{babel@language@alias@#1}{#2}%
}
\makeatother

\DeclareLanguageAlias{en}{english}

\def\DS {\scriptscriptstyle{{DS}}}
\def\B {\scriptscriptstyle{{B}}}
\def\P {\scriptscriptstyle{(P)}}
\def\Z {\scriptscriptstyle{{(Z)}}}
\def\G {\scriptscriptstyle{{G}}}
\def\F {\scriptscriptstyle{{F}}}
\def\HR {\scriptscriptstyle{{HR}}}

\usepackage{times,graphicx}
\usepackage[colorlinks=true,linkcolor=blue,urlcolor=blue,citecolor=blue]{hyperref}

\date\today
\begin{document}
	
\title{Exact Results for the Boundary Energy of One-Dimensional Bosons}
\author{Benjamin Reichert$^1$}
\author{Grigori E. Astrakharchik$^2$}
\author{Aleksandra Petkovi\'{c}$^1$}
\author{Zoran Ristivojevic$^1$}
\affiliation{$^1$Laboratoire de Physique Th\'{e}orique, Universit\'{e} de Toulouse, CNRS, UPS, 31062 Toulouse, France}
\affiliation{$^2$Departamento de F\'{i}sica, Universitat Polit\`{e}ecnica de Catalunya, Campus Nord B4-B5, 08034 Barcelona, Spain}
	
\begin{abstract}
We study bosons in a one-dimensional hard-wall box potential. In the case of contact interaction, the system is exactly solvable by the Bethe ansatz, as first shown by Gaudin in 1971. Although contained in the exact solution, the boundary energy in the thermodynamic limit for this problem is only approximately calculated by Gaudin, who found the leading order result at weak repulsion. Here we derive an exact integral equation that enables one to calculate the boundary energy in the thermodynamic limit at an arbitrary interaction. We then solve such an equation and find the asymptotic results for the boundary energy at weak and strong interactions. The analytical results obtained from the Bethe ansatz are in agreement with the ones found by other complementary methods, including quantum Monte Carlo simulations. We study the universality of the boundary energy in the regime of a small gas parameter by making a comparison with the exact solution for the hard rod gas.
\end{abstract}

\maketitle

Experimental realizations of cold gases very often involve an external confining potential to localize the atom motion in certain directions. The harmonic well is a common choice for the trapping potential \cite{kinoshita_observation_2004}. Recently, experiments with a flat box potential have been carried out in three \cite{gaunt_bose-einstein_2013,garratt_single-particle_2019}, two \cite{chomaz_emergence_2015}, and one \cite{rauer_recurrences_2018} dimension(s). The advantage of a similar shape is that it permits us to create a uniform system with hard-wall boundaries. The finite-size effects become visible, e.g., in the lowest collective excitations \cite{garratt_single-particle_2019}, which are starkly different from the behavior of the lowest frequency mode of a harmonically trapped gas, which is independent \cite{kohn_cyclotron_1961} of the interaction.
Another physical realization is a Bose gas in the presence of a single pinned impurity of infinite repulsion, which in one dimension effectively generates a similar effect to that of a hard wall. Physically, such an impurity can be a pinned atom of a different species or a laser creating a hole \cite{raman_evidence_1999} in the density.

A physical system of immense theoretical and experimental interest is the one of one-dimensional bosons with contact interaction, which is known as the Lieb-Liniger model \cite{lieb_exact_1963}. Its remarkable realizations \cite{kinoshita_observation_2004,paredes_tonks-girardeau_2004,kruger_weakly_2010,meinert_probing_2015} offer a fertile ground since many theoretical results for this model can be tested and verified with unprecedented accuracy. This includes quantum dynamics \cite{kinoshita_quantum_2006,hofferberth_non-equilibrium_2007}, solitons \cite{becker_oscillations_2008}, the crossover from the repulsive to attractive interaction regime \cite{haller_realization_2009}, quantum correlations \cite{tolra_observation_2004,armijo_probing_2010,fabbri_dynamical_2015}, etc. On the theoretical side, the Lieb-Liniger model is exactly solvable \cite{lieb_exact_1963,lieb_exact_1963b,korepin1993book} by the Bethe ansatz \cite{bethe_zur_1931}. Initially, the solution was found for periodic boundary conditions \cite{lieb_exact_1963}, but later also for zero boundary conditions \cite{gaudin_boundary_1971}. The latter case corresponds to bosons in an enclosed hard-wall box imposing the nullification of the wave function at the two systems' ends.

The case with zero boundary conditions shows some important qualitative differences. In particular, it is characterized by the boundary energy $E_{\B}$, which represents the nonextensive part of the ground-state energy $E_0$ in the thermodynamic limit \cite{gaudin_boundary_1971,blote_conformal_1986}
\begin{align}\label{eq1}
E_0=N\epsilon_0 + E_{\B}+O(1/N).
\end{align}
Here $\epsilon_0$ is the ground-state energy per particle, while $N$ is the total number of bosons. Note that the bulk energy $\epsilon_0$ is identical for the two geometries, while the boundary  energy $E_{\B}$ is a surface effect and it exists only in the case of zero boundary conditions \cite{gaudin_boundary_1971,batchelor_1d_2005}. The physical origin of $E_{\B}$ is the increase in the system energy due to the hard-wall potential, which causes the density to be nonuniform and also increases its value in the bulk region. A node in the many-body wave function at the edge leads to its nonzero gradient, increasing the kinetic energy. The typical size of the density depletion near the boundary is on the order of the healing length $\xi$ and thus involves $\xi n$ particles, where $n$ is the (mean) boson density. This enables us to estimate the boundary energy as $E_{\B}\sim \hbar^2 n/m\xi$, where $m$ denotes the mass of bosons.

The Lieb-Liniger model is characterized by two types of elementary excitations \cite{lieb_exact_1963b}. In addition to the particlelike type-I branch, the model supports holelike type-II excitations. At weak interaction, they are identified with gray soliton solutions of the mean-field Gross-Pitaevskii equation \cite{tsuzuki_nonlinear_1971,kulish_comparison_1976, ishikawa_solitons_1980,pitaevskii2018book}. A gray soliton corresponds to a localized perturbation in the boson density moving at a fixed velocity. In the case of a complete local suppression of the density, the soliton is called dark; it becomes static and its density profile is quite reminiscent of the one near the boundaries in the system with the hard-wall potential. In the weakly interacting regime, the two density deeps, around the center of the dark soliton and around the boundary, are described by the same Gross-Pitaevskii equation. Since the energy functional is local in the latter theory, the energy of the dark soliton coincides with the total boundary energy arising from the two ends, $E_{\B}$. This simple reasoning leads to the result
\begin{align}\label{eq:Ebweak}
E_{\B}=\frac{8}{3}\epsilon\sqrt{\gamma}.
\end{align}
Here $\gamma\ll 1$ is the dimensionless interaction strength defined below, while $\epsilon=\hbar^2 n^2/2m$ is the natural unit of energy for our system. We notice that at $\gamma\ll 1$ the healing length is $\xi\sim1/n\sqrt{\gamma}$ and the previous estimate of $E_{\B}$ is consistent with Eq.~(\ref{eq:Ebweak}).

In Lieb's classification \cite{lieb_exact_1963b}, the dark soliton corresponds to the type-II excitation with zero velocity, i.e., the (Fermi) momentum $\pi\hbar n$. In the limit of strong interaction, $\gamma\to\infty$, its energy can be easily found by using the dual model of free fermions \cite{girardeau_relationship_1960,yukalov_fermi-bose_2005}. The type-II excitation corresponds in the fermionic picture to the excitation where a fermion is promoted from the bottom to the top of the Fermi sea. Its energy is therefore identical to the Fermi energy, $\pi^2\epsilon$. On the other hand, the ground-state energy of $N$ free fermions in a hard-wall box of the size $L$ is $E_0=\frac{\pi^2\hbar^2}{2mL^2}\sum_{j=1}^N j^2$. Using Eq.~(\ref{eq1}) one then finds the boundary energy 
\begin{align}
E_{\B}=\frac{\pi^2}{2}\epsilon,
\end{align}
which is twice as small as the energy of the type-II excitation. The above simple arguments show that the dark soliton (i.e, the type-II excitation of the momentum $\pi\hbar n$) and the boundary energy are different, contrary to the indication that might have appeared when studying the $\gamma\ll 1$ case. 

In Ref.~\cite{gaudin_boundary_1971}, Gaudin derived the expression for the boundary energy of the Lieb-Liniger model in terms of an integral equation (see further below) that should be presumably valid at any interaction $\gamma$. However, he only solved it at weak interaction, finding the expression~(\ref{eq:Ebweak}).

In this Letter, we show that Gaudin's expression for the boundary energy actually coincides with the energy of the type-II excitation of the momentum $\pi\hbar n$ at any $\gamma$. Moreover, it differs from the exact boundary energy already at the subleading order $O(\gamma)$ in Eq.~(\ref{eq:Ebweak}). Furthermore, at strong interaction, Gaudin's expression overestimates the boundary energy two times. Instead, here we derive an exact expression for $E_{\B}$ and evaluate it analytically at strong and weak interactions. In addition, we use the Monte Carlo method as an independent check of our findings. Finally, by making a comparison with the exact solution for the gas of hard rods, we demonstrate that the behavior of the boundary energy of various systems in the regime of small densities is universal in terms of the gas parameter.

We consider bosons in one dimension described by the Lieb-Liniger Hamiltonian \cite{lieb_exact_1963,korepin1993book}
\begin{align}\label{eq:H}
H=\frac{\hbar^2}{2m}\left[-\sum_{i=1}^N\frac{\partial^2}{\partial x_i^2} + c\sum_{i\neq j}\delta(x_i-x_j)\right].
\end{align}
The local repulsion is described by the coupling constant $c$ in Eq.~(\ref{eq:H}), while the thermodynamic properties of the system are governed by the dimensionless parameter $\gamma=c/n$, where $n=N/L$ is the linear density. Here $N$ is the number of bosons and $L$ is the system size. We study the cases with periodic and zero boundary conditions corresponding, respectively, to the bosons on a ring and in a box trap.

The Hamiltonian~(\ref{eq:H}) can be diagonalized by the Bethe ansatz. The resulting equations for the ground state of a system with periodic boundary conditions of length $2L$ with $2N$ particles have the form \cite{lieb_exact_1963,korepin1993book}
\begin{align}\label{eq:BAEPBC}
2k_iL=2\pi \left(i-\frac{2N+1}{2}\right) -\sum_{j=1}^{2N} \theta(k_i-k_j),
\end{align}
where $\theta(k)=2\arctan(k/c)$ and $i=1, 2,\ldots,2N$. The system of equations~(\ref{eq:BAEPBC}) has a unique solution with distinct quasimomenta $k_i$, where one-half of them are negative ($k_i<0$ for $1\le i\le N$), while the remaining ones are positive ($k_i>0$ for $N+1\le i\le 2N$). Moreover, the quasimomenta are positioned symmetrically around zero, i.e., $k_i=-k_{2N+1-i}$. It will be convenient to shift the indices in Eq.~(\ref{eq:BAEPBC}): $i\to i-N-1$ for $1\le i\le N$ and $i\to i-N$ for $N+1\le i\le 2N$, so that one has the property $k_i=-k_{-i}$. This enables us to eventually write
\begin{align}\label{eq:BAEPBC1}
k_i L=\pi \left(i-\frac{1}{2}\right)-\frac{1}{2}\sum_{j=1}^{N}\left[ \theta(k_i-k_j) + \theta(k_i+k_j)\right],
\end{align}
where $i=1,2,\ldots, N$. The ground state of the Hamiltonian~(\ref{eq:H}) is thus characterized by the set of $N$ positive quasimomenta obtained by solving the system~(\ref{eq:BAEPBC1}), while the negative ones are automatically obtained from them. The ground-state energy is then given as
$E^{\P}(2N)=\frac{\hbar^2}{m} \sum_{i=1}^N\ k_i^2$, where the superscript denotes periodic boundary conditions.

As first shown by \citet{gaudin_boundary_1971}, the Hamiltonian~(\ref{eq:H}) can also be diagonalized for a system in a box with zero boundary conditions imposed on the wave function. The Bethe ansatz equations for the ground state in this case, for a system of length $L$ with $N$ particles, are given by \cite{gaudin_boundary_1971}
\begin{align}\label{eq:GaudinBAE}
\bar k_i L=\pi+\sum_{j=1\atop j\neq i}^{N} \left(\arctan\frac{c}{\bar k_i-\bar k_j} +\arctan\frac{c}{\bar k_i+\bar k_j}\right),
\end{align}
where $i=1,2,\ldots,N$. Equation~(\ref{eq:GaudinBAE}) allows only for $\bar k_i>0$. Using the identity $\arctan x+\arctan(1/x)=\pi\, \mathrm{sgn}(x)/2$ one can reexpress Eq.~(\ref{eq:GaudinBAE}) as
\begin{align}\label{eq:GaudinBAE1}
\bar k_i L=\pi i-\frac{1}{2}\sum_{j=1}^N\left[\theta(\bar k_i-\bar k_j)+\theta(\bar k_i+\bar k_j) \right]+\frac{\theta(2\bar k_i)}{2}.
\end{align}
The ground-state energy for this setup is given by $E^{\Z}(N)=\frac{\hbar^2}{2m}\sum_{i=1}^N \bar k_i^2$. Here the superscript denotes zero boundary conditions.

The boundary energy is the difference in the ground-state energy of the system with zero and periodic boundary conditions,
\begin{align}\label{eq:Ebdefinition}
E_{\B}(N)=E^{\Z}(N)-E^{\P}(N).
\end{align}
For the latter case, one can show that, at the same density, the energy of the systems with $N$ and $2N$ particles are simply related as $E^{\P}(N)=E^{\P}(2N)/2+O(1/N)$ \cite{gaudin_boundary_1971}. In the thermodynamic limit this yields $E_{\B}=\lim_{N\to\infty} \left[E^{\Z}(N)-E^{\P}(2N)/2\right]$, i.e.,
\begin{align}\label{eq:Eb}
E_{\B}=\lim_{N\to\infty}\frac{\hbar^2}{2m}\sum_{i=1}^N (\bar k_i^2-k_i^2),
\end{align}
where the corresponding quasimomenta are the solutions of Eqs.~(\ref{eq:GaudinBAE1}) and (\ref{eq:BAEPBC1}).

For the evaluation of the boundary energy~(\ref{eq:Eb}) we subtract Eq.~(\ref{eq:BAEPBC1}) from Eq.~(\ref{eq:GaudinBAE1}). Since in a long system the difference $\bar k_i-k_i=\Delta k_i=O(1/L)$ is small, we obtain
\begin{align}\label{eq:BE1}
\Delta k_i L={}&\frac{\pi}{2}+\frac{\theta(2\bar k_i)}{2} - \frac{1}{2} \sum_{j=1}^N [\theta'(k_i-k_j)(\Delta k_i-\Delta k_j)\notag\\
&\theta'(k_i+k_j)(\Delta k_i+\Delta k_j)]+O(1/N).
\end{align}
In a system of length $2L$ with periodic boundary conditions we define the density of quasimomenta as $\rho(k_i)=[2L(k_{i+1}-k_i)]^{-1}$. In the thermodynamic limit it satisfies the Lieb integral equation \cite{lieb_exact_1963,korepin1993book}
\begin{align}\label{eq:rho}
\rho(k)-\frac{c}{\pi}\int_{-Q}^{Q} \frac{dk' \rho(k')}{c^2+(k'-k)^2}=\frac{1}{2\pi}.
\end{align}
Here the Fermi rapidity $Q$ is fixed by the normalization condition $n=\int_{-Q}^{Q}\rho(k) dk$. Using the formal expression $\rho(k)=\sum_{i=1}^N [\delta(k-k_i)+\delta(k+k_i)]/2L$ and the property $\rho(k)=\rho(-k)$, we then obtain
\begin{align}
1+\frac{1}{2L}\sum_{j=1}^N [\theta'(k-k_j)+\theta'(k+k_j)]=2\pi \rho(k).
\end{align}
The latter equation enables us to simplify Eq.~(\ref{eq:BE1}). Introducing an odd function $g(k_i)=L\rho(k_i)\Delta k_i$, we obtain that it satisfies an integral equation
\begin{subequations}\label{eq:main}
\begin{gather}\label{eq:integralequationforg}
g(k)-\frac{c}{\pi}\int_{-Q}^{Q} \frac{dk' g(k')}{c^2+(k'-k)^2}=r(k),\\
r(k)=\frac{\mathrm{sgn(k)}}{4}+ \frac{\arctan\frac{2k}{c}}{2\pi}.\label{eq:r}
\end{gather}
\end{subequations}
The boundary energy can then be expressed as
\begin{align}\label{eq:Ebfinal}
E_{\B}=\frac{\hbar^2}{m}\int_{-Q}^Q k g(k) dk.
\end{align}
Equation~(\ref{eq:main}) is our main results. Together with Eq.~(\ref{eq:Ebfinal}) they establish the exact result for the boundary energy of the Lieb-Liniger model at an arbitrary interaction strength $c>0$.

To analyze the boundary energy, let us introduce Green's function for the Lieb integral equation as \cite{takahashi}
\begin{align}\label{eq:G}
G(k,k')-\frac{c}{\pi}\int_{-Q}^{Q} \frac{dk'' G(k',k'')}{c^2+(k-k'')^2}=\delta(k-k').
\end{align}
One can show by the method of iterations that Green's function is symmetric, $G(k,k')=G(k',k)$. Multiplying Eq.~(\ref{eq:G}) by $r(k')$ [see Eq.~(\ref{eq:r})] and performing the integration over $k'$, one obtains the integral equation~(\ref{eq:integralequationforg}) provided $g(k)=\int_{-Q}^{Q}d k' G(k,k') r(k')$. The boundary energy~(\ref{eq:Ebfinal}) then acquires the form
\begin{align}\label{eq:Ebsigma}
E_{\B}=\int_{-Q}^{Q}dk\sigma(k) r(k),
\end{align}
where we have defined $\sigma(k)=(\hbar^2/m)\int_{-Q}^{Q}d k' k' G(k,k')$. From Eq.~(\ref{eq:G}) one finds that $\sigma(k)$ satisfies
\begin{align}\label{eq:sigma}
\sigma(k)-\frac{c}{\pi}\int_{-Q}^{Q} \frac{dk' \sigma(k')}{c^2+(k-k')^2}=\frac{\hbar^2}{m}k.
\end{align}
We have therefore reformulated the problem of finding  the boundary energy to be the equivalent, but more convenient, problem of solving Eq.~(\ref{eq:sigma}) and then evaluating $E_{\B}$ of Eq.~(\ref{eq:Ebsigma}).

\begin{figure}
\centering
\includegraphics[width=\columnwidth]{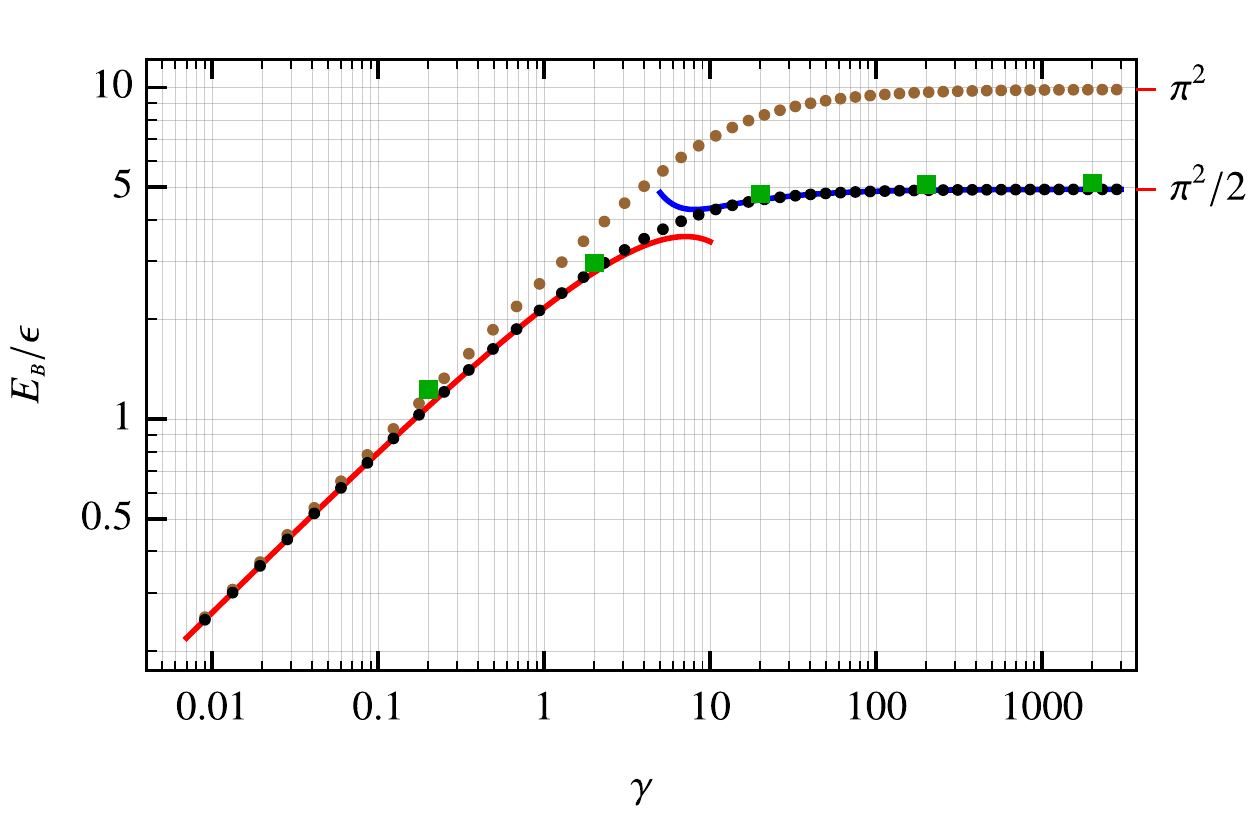}
\caption{The boundary energy $E_{\B}$ in units of $\epsilon$ as a function of the interaction strength $\gamma$. The lower (black) dots represent the exact numerically obtained results, while the two asymptotic behaviors at small and large $\gamma$ are given by formulas~(\ref{eq:Ebsmall}) and (\ref{eq:Eblarge}). The upper (brown) dots represent the result of Gaudin \cite{gaudin_boundary_1971} and coincides with the energy of Lieb's type-II excitation with zero velocity (momentum $\pi \hbar n$) in the model with periodic boundary conditions. The (green) rectangles represent the boundary energy obtained from the Monte Carlo method for $N=41$ particles, which approach the exact curve with increasing $N$.}\label{fig:plot}
\end{figure}

Additional analytical results can be obtained in the Gross-Pitaevskii and Tonks-Girardeau regimes of weak~($\gamma\ll 1$) and strong~($\gamma\gg 1$) interactions, respectively. In the former case, the integral equation for the density~(\ref{eq:rho}) is solved to first two orders in Refs.~\cite{hutson_circular_1963,popov_theory_1977}, enabling us to express $Q$ in terms of $\gamma$. However, for the boundary energy we have to solve Eq.~(\ref{eq:sigma}) within the same accuracy \footnote{See Supplemental Material for the details of calculation.}. Using Eq.~(\ref{eq:Ebsigma}) we then find
\begin{align}\label{eq:Ebsmall}
E_{\B}=\frac{8}{3}\epsilon\sqrt{\gamma}\left[1-\frac{3}{16}\sqrt{\gamma}+O(\gamma) \right],
\end{align}
which agrees at the leading order with the result~(\ref{eq:Ebweak}). In the opposite regime of strong interaction, the integral equations~(\ref{eq:rho}) and~(\ref{eq:sigma}) can be perturbatively solved by iterations to an arbitrary order in $1/\gamma$ \cite{ristivojevic_excitation_2014}. It yields \cite{Note1}
\begin{align}\label{eq:Eblarge}
E_{\B}=\frac{\pi^2}{2}\epsilon\left[1-\frac{4}{3\gamma}-\frac{4}{3\gamma^2}+\frac{4(120+7\pi^2)}{15\gamma^3}+O\left(\gamma^{-4}\right) \right].
\end{align}
In Fig.~\ref{fig:plot} we show the two asymptotic expressions and the exact data obtained by numerically evaluating Eq.~(\ref{eq:Ebfinal}) or, equivalently, Eq.~(\ref{eq:Ebsigma}).

In Ref.~\cite{gaudin_boundary_1971}, Gaudin found the integral equation of the form~(\ref{eq:integralequationforg}) but with a different right-hand side, which instead was given by $r_{\G}(k)=\mathrm{sgn(k)}/2$. Such expression is approximately the correct right-hand side of Eq.~(\ref{eq:integralequationforg}) only at $c\to 0$, as one can see by considering Eq.~(\ref{eq:r}) in this limit. Thus, Gaudin was able only to find the leading order expression~(\ref{eq:Ebweak}) for the boundary energy at weak interaction. We notice that Gaudin's result for $r_{\G}(k)$ leads to a significant overestimation of the boundary energy, see Fig.~\ref{fig:plot}. Interestingly, using Eq.~(\ref{eq:Ebsigma}) Gaudin's formula for the boundary energy becomes $E_{\B,\G}=\int_0^Qdk \sigma(k)$. Such expression formally coincides with the energy of Lieb's type-II excitation in the (periodic) Lieb-Liniger model with the momentum $\pi\hbar n$ \cite{korepin1993book,pustilnik_low-energy_2014,petkovic_spectrum_2018}. The asymptotic form of $E_{\B,\G}$ in the two regimes is given by \cite{Note1}
\begin{align}\label{eq:Ebg}
E_{\B,\G}=\epsilon
\begin{cases}
\frac{8}{3}\sqrt{\gamma}\left[1-0\cdot\sqrt\gamma + O(\gamma)\right],\\
\pi^2\left[1-\frac{4}{\gamma}+\frac{12}{\gamma^2}+\frac{4(\pi^2-8)}{\gamma^3}+O(\gamma^{-4})\right].
\end{cases}
\end{align}
At weak interaction, $E_{\B,\G}$ of Eq.~(\ref{eq:Ebg}) and $E_{\B}$ of Eq.~(\ref{eq:Ebsmall}) differ at the subleading $O(\gamma)$ order. In other words, already in the first beyond mean-field correction to the energy, there is a difference between the dark soliton and the boundary energy. At large $\gamma$, $E_{\B,\G}$ is twice $E_{\B}$  (see Fig.~\ref{fig:plot}).

Additional physical insights for the boundary energy can be obtained by using more elementary approaches than the Bethe ansatz. The weakly interacting case $\gamma\ll 1$ can be studied using the Gross-Pitaevskii equation and the quantum corrections to it. Such procedure indeed recovers the boundary energy~(\ref{eq:Ebsmall}) \cite{reichert_fluctuation-induced_2019}. In the opposite regime of strong interaction between bosons $\gamma\gg 1$, one can study the model~(\ref{eq:H}) using the perturbation theory on the related dual Cheon-Shigehara model of fermions of the same mass $m$, which interact via the attractive potential $V_{\F}(x)=-(2\hbar^2/m c)\delta''(x)$ \cite{cheon_fermion-boson_1999,sen_fermionic_2003,yukalov_fermi-bose_2005,khodas_dynamics_2007}. In the noninteracting limit of fermions \cite{girardeau_relationship_1960} in a box one obtains the boundary energy $\pi^2\epsilon/2$, while the linear correction in $V_{\F}$ reproduces the first correction $\propto1/\gamma$ of Eq.~(\ref{eq:Eblarge}) \cite{Note1}.

We also calculated the boundary energy using the diffusion Monte Carlo method. In this approach one approximates the many-body wave function by the product
$\psi(x_1,x_2,\ldots,x_N) = \prod_{i=1}^{N}f_1(x_i)\prod_{i<j}^{N}f_2(x_i-x_j)$. The one-body term is chosen as $f_1(x) = \sin^\alpha(\pi x/L)$ and it imposes the zero boundary conditions. The remaining two-body Jastrow terms are constructed \cite{astrakharchik_correlation_2003,astrakharchik_beyond_2005,petrov_ultradilute_2016,parisi_spin_2018,parisi_liquid_2019} at short distances from the two-body scattering solution, $f_2(x) = C_1\cos(k(|x|-C_2)), |x|<C_3$, which satisfies the Bethe-Peierls boundary condition and from the phononic tail at larger distances \cite{reatto_phonons_1967},
$f_2(x) = \sin^{1/K}(\pi|x|/L), |x|>C_3$, where $K$ is the Luttinger liquid parameter. The free parameter $\alpha$ is fixed by minimizing the variational energy, $K$ is taken from the Bethe ansatz solution \cite{lieb_exact_1963}, while the constants $C_1, C_2$, and $C_3$ are fixed by the boundary and the continuity conditions.

The diffusion Monte Carlo method is used to obtain the boundary energy at several values of $\gamma$ for $N=21$ and $N=41$ particles. Both sets of results are in agreement with the boundary energy obtained by numerically solving the discrete Bethe ansatz equations. The boundary energy for $N=21$ particles is always slightly larger than the one for $N=41$, which approaches the exact value of $E_{\B}$ in the thermodynamic limit, see Fig.~\ref{fig:plot}. The results for $N=21$ are not shown because they would be hardly distinguishable from the ones of $N=41$ on the resolution of Fig.~\ref{fig:plot}.

In the limit of low density, specific details of short-range potentials become irrelevant and a single parameter, namely the $s$-wave scattering length $a$, is sufficient to represent the potential. In order to verify the universality of the boundary energy in terms of the gas parameter $na$, we consider a gas of hard rods with the diameter $a>0$. As noted by \citet{girardeau_relationship_1960}, the wave function and the energy of such gas can be obtained from the Tonks-Girardeau gas by subtracting the excluded volume as the total accessible volume of the phase space is reduced by $Na$ in the case of periodic boundary conditions and by $(N-1)a$ for zero boundary conditions. The difference in the reduced space arises from the physical difference between particles on a ring (for example, a single particle interacts with its own image) and zero boundary condition. In the thermodynamic limit, we find the boundary energy of hard rods to be
\begin{align}\label{eq:EbHR}
E_{\B}^{\HR}
=\frac{\pi^2}{2}\epsilon \frac{1+\frac{14}{3\gamma}}{\left(1+\frac{2}{\gamma}\right)^3}
=\frac{\pi^2}{2}\epsilon
\left[1-\frac{4}{3\gamma}-\frac{4}{\gamma^2}+O\left(\gamma^{-3}\right) \right],
\end{align}
where $\gamma = -2/na<0$. By comparison with Eq.~(\ref{eq:Eblarge}) derived for delta-interacting gas and $\gamma>0$, one finds that the first two terms are universal. This provides the physical interpretation of the leading terms as arising from the excluded-volume effect. The validity of the excluded-volume correction to the Lieb-Liniger gas has been verified in Ref.~\cite{astrakharchik_low-dimensional_2010} for the ground state and in Ref.~\cite{de_rosi_beyond-luttinger-liquid_2019} for the thermal (Yang-Yang) state. Another relevant consequence is that the boundary energy expressed in terms of the gas parameter is expected to be universal in rather different physical systems, including the gases with dipolar \cite{Arkhipov2005,citro_evidence_2007,girardeau_super-tonks-girardeau_2012} and Rydberg \cite{osychenko_phase_2011} interactions as well as for bosonic $^4\mathrm{He}$ \cite{bertaina_one-dimensional_2016} and fermionic $^3\mathrm{He}$ \cite{astrakharchik_luttinger-liquid_2014} in the regime of low densities. Alternatively, the boundary energy in an excited super Tonks-Girardeau gas \cite{astrakharchik_beyond_2005,Haller2009} will follow Eq.~(\ref{eq:EbHR}) at small densities. However, $\gamma$ is negative in this case and thus the boundary energy will be larger in comparison to the Tonks-Girardeau limit.

Let us finally notice that the boundary energy~(\ref{eq:Ebfinal}) is derived in the thermodynamic limit, when the system size is much larger than the healing length, $L\gg\xi$. In a finite system there is an additional regime where $L\lesssim \xi$, which can occur only at very weak interaction that satisfies $\gamma\lesssim 1/N^2$. We leave this problem for a future study. We also notice that Eq.~(\ref{eq1}) has finite-size corrections \cite{blote_conformal_1986} that vanish in the thermodynamic limit.

In conclusion, we have found the exact results for the boundary energy of the experimentally relevant Lieb-Liniger model. We derived the governing integral equation that we analytically solved in the regimes of weak and strong interaction, while numerically we solved it everywhere. We showed that in the initial work \cite{gaudin_boundary_1971} and the book \cite{gaudin_2014} of Gaudin, the boundary energy was actually coincident with the energy of the type-II excitation with the momentum $\pi\hbar n$. The latter excitation, which at weak interaction becomes the dark soliton, has always a greater energy than the true boundary energy at any repulsion, see Fig.~\ref{fig:plot}. Our Letter thus corrects the old misconception, making a clear distinction between the dark soliton and the boundary energy of the Lieb-Liniger model.

G.~E.~A.~acknowledges useful discussions with L.~P.~Pitaevskii and V.~A.~Yurovky. This study has been partially supported through the EUR Grant No.~NanoX ANR-17-EURE-0009 in the framework of the ``Programme des Investissements d’Avenir." G.~E.~A.~acknowledges funding from the Spanish MINECO (FIS2017-84114-C2-1-P).
The Barcelona Supercomputing Center (The Spanish National Supercomputing Center - Centro Nacional de Supercomputaci\'on) is acknowledged for the provided computational facilities (RES-FI-2019-2-0033).


%


\onecolumngrid
\newpage
\setcounter{equation}{0}
\setcounter{figure}{0}

\renewcommand{\theequation}{S\arabic{equation}}
\renewcommand{\thepage}{S\arabic{page}}
\renewcommand{\thesection}{S\arabic{section}}
\renewcommand{\thetable}{S\arabic{table}}
\renewcommand{\thefigure}{S\arabic{figure}}

\renewcommand{\bibnumfmt}[1]{[{\normalfont S#1}]}



\begin{center}
	{\large\textbf{Exact Results for the Boundary Energy of One-Dimensional Bosons}
		\\\vskip 5pt
		\normalsize{--Supplemental material--}
		\\\vskip 5pt
	}
	
	Benjamin Reichert$^1$, Grigori E. Astrakharchik$^2$, Aleksandra Petkovi\'{c}$^1$, and Zoran Ristivojevic$^1$	\vskip 0.5mm
	\textit{\small$^1$Laboratoire de Physique Th\'{e}orique, Universit\'{e} de Toulouse, CNRS, UPS, 31062 Toulouse, France}\\
	\textit{\small $^2$Departamento de F\'{i}sica, Universitat Polit\`{e}ecnica de Catalunya, Campus Nord B4-B5, 08034 Barcelona, Spain}
	
\end{center}
\vskip 0mm

\section{Bethe ansatz equations}

If we introduce the dimensionless units and rescale all momenta by $Q$, the set of Bethe ansatz equations for $\rho$ and $\sigma$ of the main text, respectively, become
\begin{gather}\label{eq1}
\varrho(x)-\dfrac{\lambda}{\pi}\int_{-1}^1dx'\dfrac{\varrho(x')}{\lambda^2+(x-x')^2} =\dfrac{1}{2\pi},\\
\label{eq2}
\varsigma(x)-\dfrac{\lambda}{\pi}\int_{-1}^1dx'\dfrac{\varsigma(x')}{\lambda^2+(x-x')^2} = x.
\end{gather}
The normalization condition is then reexpressed as
\begin{align}
\gamma \int_{-1}^1dx\varrho(x)=\lambda.
\end{align}
We notice that $\lambda=c/Q$. The boundary energy $E_{\B}$ and the energy $E_{\DS}$ of type~II excitation with the momentum $\pi\hbar n$ are given by
\begin{gather}
E_{\B}=2\epsilon \frac{\gamma^2}{\lambda^2}\int_0^1 dx\varsigma(x)\left(\dfrac{1}{2}+\dfrac{1}{\pi}\arctan\frac{2x}{\lambda} \right),\\
E_{\DS}=E_{\B,\G}=2\epsilon \frac{\gamma^2}{\lambda^2}\int_0^1 dx\varsigma(x),\quad \epsilon=\frac{\hbar^2 n^2}{2m}.
\end{gather}

\subsection{Weakly interacting limit}

The solution of Eq.~(\ref{eq1}) to the first two orders is the regime of weak interaction was found by Popov \cite{popov_theory_1977}:
\begin{align}\label{eq0}
\varrho(x)=\frac{\sqrt{1-x^2}}{2\pi\lambda} +\frac{1+\ln\left(\frac{16\pi}{\lambda}\right) -x\ln\frac{1+x}{1-x}}{4\pi^2\sqrt{1-x^2}}+O(\lambda).
\end{align}
Equation (\ref{eq0}) applies for $x$ no too close to Fermi rapidities, i.e., it is valid at $1-x^2\gg\lambda$. However for our purpose this limitation turns out not to be important and thus we will integrate $\varrho(x)$ from $-1<x<1$. This leads to
\begin{align}
\lambda=\dfrac{\sqrt{\gamma}}{2}-\gamma\frac{1-\ln\left(\frac{ 32\pi}{\sqrt\gamma}\right)}{8\pi}+O(\gamma^{3/2}).
\end{align}
Using the approach of Ref.~\cite{popov_theory_1977}, we solved Eq.~(\ref{eq2}) within the same accuracy. We found
\begin{gather}
\varsigma(x)=\frac{x\sqrt{1-x^2}}{2\lambda}+\dfrac{x\left( 1+\ln\frac{16\pi}{\lambda}\right)+(1-2x^2)\ln\frac{1+x}{1-x}}{4\pi\sqrt{1-x^2}}+ O(\lambda).
\end{gather}
Notice that the same comment for the range of $x$ as above for $\varrho(x)$ applies for $\varsigma(x)$. We eventually obtain
\begin{gather}
E_{\B}=\frac{8}{3}\epsilon\sqrt{\gamma}\left[1-\frac{3\sqrt\gamma}{16} + O(\gamma)\right],\label{wlg}\\
E_{\DS}=\frac{8}{3}\epsilon\sqrt{\gamma}\left[1-0\cdot\sqrt\gamma + O(\gamma)\right].
\end{gather}
Therefore, at weak interaction the boundary energy differs from the energy of the dark soliton at the subleading $O(\gamma)$ order.

\subsection{Strongly interacting limit}
In the strongly interacting limit the integral equations (\ref{eq1})--(\ref{eq2}) are systematically solved in Ref.~\cite{ristivojevic_excitation_2014}, yielding
\begin{gather}
\varsigma(x)=x\left(1+\dfrac{4}{3\pi\lambda^3}\right)+O(\lambda^{-5}),\\
\lambda=\dfrac{\gamma}{\pi}+\dfrac{\pi}{2}-\dfrac{4\pi}{3\gamma^2}+\dfrac{16\pi}{3\gamma^3}+O(\gamma^{-4}).
\end{gather}
This leads to
\begin{gather}
E_{\B}=\frac{\pi^2}{2}\epsilon \left[1-\dfrac{4}{3\gamma}-\dfrac{4}{\gamma^2}+\dfrac{4(120+7\pi^2)}{15\gamma^3} -\dfrac{40\left(30+\pi^2\right)}{9\gamma^4}+ O(\gamma^{-5})\right],\\
E_{\DS}=\pi^2\epsilon\left[1-\frac{4}{\gamma}+\frac{12}{\gamma^2}+\frac{4(\pi^2-8)}{\gamma^3}-\frac{40(\pi^2-2)}{\gamma^4}+O(\gamma^{-5})\right].
\end{gather}
The leading correction to $E_{\B}$ is in agreement with the calculation within a dual fermionic model, as we demonstrate below.

\section{Perturbation theory for strongly interacting bosons \label{appB}}

We study the strongly interacting limit of the Lieb-Liniger model using the dual Cheon-Shigehara model. It is characterized by the two-particle interaction
\begin{align}\label{seV}
V(x)=\lambda\delta''(x),\quad \lambda=-\dfrac{2\hbar^2}{m c}.
\end{align}
The fermions have the same mass as bosons, $m$. Equation (\ref{seV}) shows that the strong repulsion between the bosons, $c\gg n$, corresponds to the weak attraction between fermions, which is convenient as one can calculate the ground-state energy using perturbation theory. The Hamiltonian of $N$ weekly interacting fermions in a box of size $L$ is $H=H_0+H_I$ where
\begin{gather}
H_0=\frac{\hbar^2}{2m}\int_0^L dx(\nabla\psi^\dagger )(\nabla\psi),\\
H_I=\frac{1}{2}\int_0^L dx dy \psi^\dagger(x)\psi^\dagger(y) V(x-y) \psi(y)\psi(x).
\end{gather}
Here $\psi$ is the single particle operator for fermions of the mass $m$ with the standard anti-commutation relations $\{\psi(x),\psi^\dagger(y)\}=\delta(x-y)$ and $\{\psi(x),\psi(y)\}=0$.

In a box of size $L$ with the hard wall boundary conditions, the single particle operators take the form
\begin{align}
\psi(x)=\sqrt{\frac{2}{L}}\sum_{k>0} \sin(k x)a_k, \quad
\psi^\dagger(x)=\sqrt{\frac{2}{L}}\sum_{k>0} \sin(k x)a^\dagger_k,
\end{align}
where $k$ is quantized as $k=\pi j/L$. Here $j$ is a positive integer.
The kinetic energy then becomes
\begin{align}
H_0=\sum_{k>0}\dfrac{\hbar^2k^2}{2m}a_k^\dagger a_k,
\end{align}
while the interaction is given by
\begin{align}
H_I={}&\frac{\lambda}{4L}\sum_{k_1,..,k_4>0}a_{k_1}^\dagger a_{k_2}^\dagger a_{k_3} a_{k_4} \left[(k_2+k_3)^2(\delta_{k_1,k_2+k_3+k_4}+\delta_{k_4,k_1+k_2+k_3} -\delta_{k_1+k_4,k_2+k_3}) \right. \notag\\
& +(k_2-k_3)^2(\delta_{k_3,k_1+k_2+k_4} -\delta_{k_1+k_2,k_3+k_4}+\delta_{k_2,k_1+k_3+k_4}-\delta_{k_1+k_3,k_2+k_4})].
\end{align}

In the framework of perturbation theory, the ground-state energy is given by
\begin{align}
E=\langle \Omega|H_0|\Omega\rangle +\langle \Omega | H_I |\Omega \rangle+\cdots,
\end{align}
where the filled Fermi sea is
\begin{align}
|\Omega\rangle=\left(\prod_{i=1}^{N} a_{\pi i/L}^\dagger\right)|0\rangle.
\end{align}
Here $|0\rangle$ denotes the vacuum. We notice the property
\begin{align}
\langle \Omega| a_k^\dagger a_q|\Omega\rangle=\delta_{k,q} \theta_H(k_F-k),
\end{align}
where $k_F=\pi N/L$ and $\theta_H$ is the Heaviside step function. We then obtain the average kinetic energy
\begin{align}
\langle \Omega|H_0|\Omega\rangle={}&\frac{\hbar^2 }{2m}\frac{\pi^2 N(1+N)(2N+1)}{6L^2}\notag\\
={}&N\epsilon\left[\frac{\pi^2}{3}+\frac{\pi^2}{2N}+O(N^{-2})\right],\quad \epsilon=\frac{\hbar^2 n^2}{2m}.
\end{align}
The leading interaction correction to it is
\begin{align}
\langle \Omega | H_I |\Omega \rangle={}&\frac{\pi^2}{6L^2}\lambda n (N^2-1)(2N+1)\notag\\
={}&-\dfrac{1}{\gamma}N\epsilon \left[\dfrac{4\pi^2}{3}+\dfrac{2\pi^2}{3N}+O(N^{-2}) \right].
\end{align}
If we express the ground-state energy as $E=N\epsilon_0+E_{\B}+O(1/N)$, we find
\begin{gather}
\epsilon_0=\frac{\pi^2}{3} \epsilon \left[1-\dfrac{4}{\gamma}+ O(\gamma^{-2})\right],\\
E_{\B}=\frac{\pi^2}{2} \epsilon \left[1-\dfrac{4}{3\gamma}+O(\gamma^{-2})\right],
\end{gather}
which is in agreement with the Bethe ansatz calculation.
\end{document}